\documentclass[]{aastex631}

\def\gangli#1{{\color{black}{#1}}}

\graphicspath{{./}{Figures/}}

\begin{document}

\title{Modelling Solar Energetic Neutral Atoms from Solar Flares and CME-driven Shocks}

\author[0000-0003-4695-8866]{Gang Li}
\affiliation{Department of Space Science and CSPAR \\
University of Alabama in Huntsville \\
Huntsville, AL 35899, USA}

\author[0000-0001-6874-2594]{Albert Y. Shih}
\affiliation{NASA Goddard Space Flight Center, Greenbelt, MD 20771, USA}

\author[0000-0003-2079-5683]{Robert C. Allen}
\affiliation{Johns Hopkins University Applied Physics Laboratory, Laurel, MD, 20723, USA}

\author[0000-0003-1093-2066]{George C. Ho}
\affiliation{Johns Hopkins University Applied Physics Laboratory, Laurel, MD, 20723, USA}

\author[0000-0002-0978-8127]{Christina M.S. Cohen}
\affiliation{California Institute of Technology, Pasadena, CA, 91125, USA}

\author[0000-0002-7318-6008]{Mihir Desai}
\affiliation{Southwest Research Institute, San Antonio, TX, 78238, USA}

\author[0000-0001-9323-1200]{Maher A. Dayeh}
\affiliation{Southwest Research Institute, San Antonio, TX, 78238, USA}

\author[0000-0003-2169-9618]{Glenn Mason}
\affiliation{Johns Hopkins University Applied Physics Laboratory, Laurel, MD, 20723, USA}



\begin{abstract}
We examine the production of energetic neutral atoms (ENAs) in solar flares and CME-driven shocks and their subsequent propagation to 1 au. Time profiles and fluence spectra of solar ENAs at 1 au are computed for two scenarios: 1) ENAs are produced downstream at CME-driven shocks, and 2) ENAs are produced at large-scale post-flare loops in solar flares. Both the time profiles and fluence spectra  for these two scenarios are vastly different. Our calculations indicate that we can use solar ENAs as a new probe to examine the underlying acceleration process of solar energetic particles (SEPs)  and to differentiate the two acceleration sites: large loops in solar flares and downstream of  CME-driven shocks, in large SEP events. \end{abstract}




\section{Introduction} \label{sec:intro}


Solar flares and coronal mass ejections (CMEs) are two of the most energetic processes in the solar system.  Efficient particle acceleration can 
occur in both solar flares and at CME-driven shocks. 
Energetic protons accelerated at either CME-driven shocks or solar flares can precipitate down to the Sun’s surface or propagate into the interplanetary medium along open interplanetary magnetic field (IMF) lines. During their propagation, they can interact with ions and thermal neutral atoms in the solar atmosphere via charge exchange, and produce energetic neutral hydrogen atoms. 
Once produced, energetic neutral hydrogen atoms (hereafter referred as ENAs) do not feel solar magnetic field and propagate along straight lines. They are subject to loss processes wherein they lose the electron and become an energetic proton again. Because the density of the solar wind drops quickly with the heliocentric distance, and because the loss rate of ENAs is proportional to the solar wind density, ENAs reaching 20 Rs suffer no further loss.  Since the IMF does not affect the propagation of ENA hydrogen, these ENAs therefore provide a powerful avenue in probing the acceleration processes and plasma properties of the underlying acceleration site.

Because the production cross section is small, the flux of ENAs at a distance of 1 au from the Sun can be extremely small.  To date, only a few observational clues of ENAs accompanying SEP events were reported 
\citep{Mewaldt2009ApJ...693L..11M, Mason2021ApJ...923..195M}. 
\citet{Mewaldt2009ApJ...693L..11M} reported $1.6$ to $5$ MeV energetic neutral atoms (ENAs) from STEREO-A/B observations. They inferred a power-law spectrum of $dJ/dE \sim  E^{-2.46}$ accompanying an X9-class solar flare and suggested that these ENAs are produced via charge exchange of SEP protons with O$^{6+}$ ions.
Following \citep{Mewaldt2009ApJ...693L..11M}, \citet{Wang2014ApJ...793L..37W} performed a simulation and showed that sufficient counts of ENAs are expected for typical gradual SEP events where particles are accelerated at CME-driven shocks. This stimulated interests in observational effort. 
More recently \citet{Mason2021ApJ...923..195M} examined $18$ SEP events with SAMPEX, and found indirect, but compelling evidence of solar ENAs near the geomagnetic equator at low altitudes where the geomagnetic field filters out all charged SEPs. This new insight also shed light on three previously reported puzzling $\sim$ MeV ion intensity increases that were also observed near the equatorial regions about  $\sim 3$ hrs after the occurrence of the corresponding X-ray flares \citep{Greenspan1999JGR...10419911G}. The discovery of ENAs by STEREO, and confirmation from SAMPEX, shows that solar ENAs can be expected to accompany many large SEP events. 

\begin{figure}[ht]
    \centering
   \includegraphics[width=0.5\textwidth]{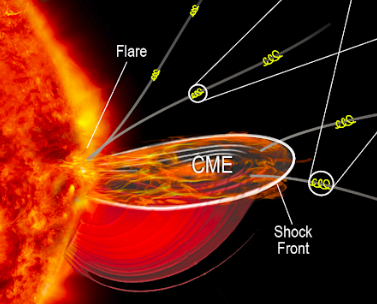}
    \caption{Schematic cartoon showing the two acceleration sites of ions in large SEP events: CME-driven shocks and flares. Once produced, energetic ions can propagate along open IMFs and be detected in-situ at 1 au. Near the acceleration sites, the density of solar atmosphere is high enough so that energetic ions can lead to the production of solar ENAs. The characteristics of solar ENAs in both scenarios are examined in this work.
      } \label{fig:accSites}
\end{figure}

If ENAs can be detected in SEP events, one of the pressing questions would be where do they originate. Are they accelerated at a rather confined reconnection site at flares or at a broader shock front driven by CMEs?  To answer such a question, we examine ENA productions in two different scenarios: CME-driven shock and large post flare loops in this work.  A schematic of the two acceleration sites are shown in Figure~\ref{fig:accSites}. Note that in many large SEP events, CMEs and flares often occur together.  
However, the spatial extension of the flare is much smaller than the CME. Ions can be efficiently accelerated at both the flare site and the CME-driven shock front. In the case of CME-driven shocks, protons and ions are accelerated at the shock front via the first order Fermi acceleration mechanism. Once accelerated, they can escape upstream propagating along IMF, or trapped downstream for an extended period of time. They may precipitate down to the solar surface, causing, for example, long duration gamma-ray events \citep{Share018ApJ...869..182S,JinMeng2018ApJ...867..122J}. 
In the case of flares, particles can be accelerated at the reconnection exhausts and in solar flare loops \citep{Petrosian2012SSRv..173..535P,Ryan2000SSRv...93..581R} by e.g. the second order Fermi acceleration mechanism. Continued magnetic reconnection can lead to a rising of the post-flare loops \citep{West2015ApJ...801L...6W}. Accelerated particles may be trapped in post-flare loops for very long period of time, serving as an alternative candidate for the long-duration gamma ray events \citep{Ryan2000SSRv...93..581R, deNolfo2019ApJ...879...90D}.

\gangli{
We note that in large SEP events it is possible that flare accelerated particles can be re-accelerated at the accompanying CME-driven shocks \citep{Li2005MixedPtcl, Petrosian2012SSRv..173..535P}.  Simulations by \citet{Li2005MixedPtcl} showed that depending on if the observer is magnetically connected to the flare and/or the shock surface, the characteristics of the ion time profiles differ and may show two peaks as reported in  \citep{Cane2003GeoRL..30.8017C}. However, because the presence of solar wind MHD turbulence can affect the propagation of charged ions, the interpretation of 1-au ion observations is often complicated. ENAs, with ballistic propagation, do not follow IMF and are not affected by the solar wind MHD turbulence. Therefore, ENA observations with enough angular resolution can clearly distinguish ENAs from flare sites and those from much broader CME-driven shocks. 
}

We examine solar ENA production from CME-dirven shocks in section~\ref{sec:CmeENAs} and from solar flare loops in section~\ref{sec:FlareENAs}. 
Production and loss processes of ENAs are discussed in Appendix~\ref{sec:ProductionLoss}.

\section{ENAs from CME-driven shocks} \label{sec:CmeENAs}

In this section, we consider the observation of ENA particles generated at a propagating CME-driven shock. The first ENA simulation was done by \citet{Wang2014ApJ...793L..37W} who simulated a CME-driven shock from a side-on orientation and suggested that the observed flux in \citep{Mewaldt2009ApJ...693L..11M} is consistent with ENA production at a CME-driven shock.
More recently, following the work of \citep{Wang2014ApJ...793L..37W}, 
\citet{Wang2022E&PP....6...42W} examined a variety cases with different CME speeds, open angles, and CME propagation directions. They also examined the effect of solar wind density variation near the Sun on the production of ENAs. These authors found similar results as \citet{Wang2014ApJ...793L..37W}.

\begin{figure}[ht]
    \centering
     \includegraphics[width=0.47\textwidth]{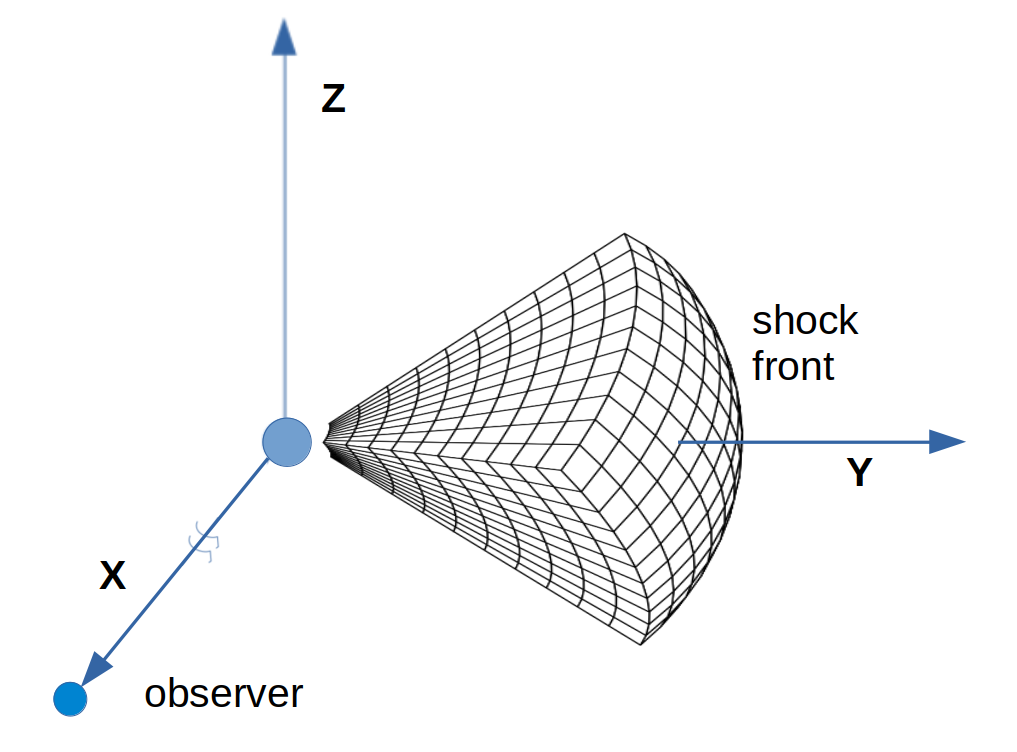}
   \includegraphics[width=0.48\textwidth]{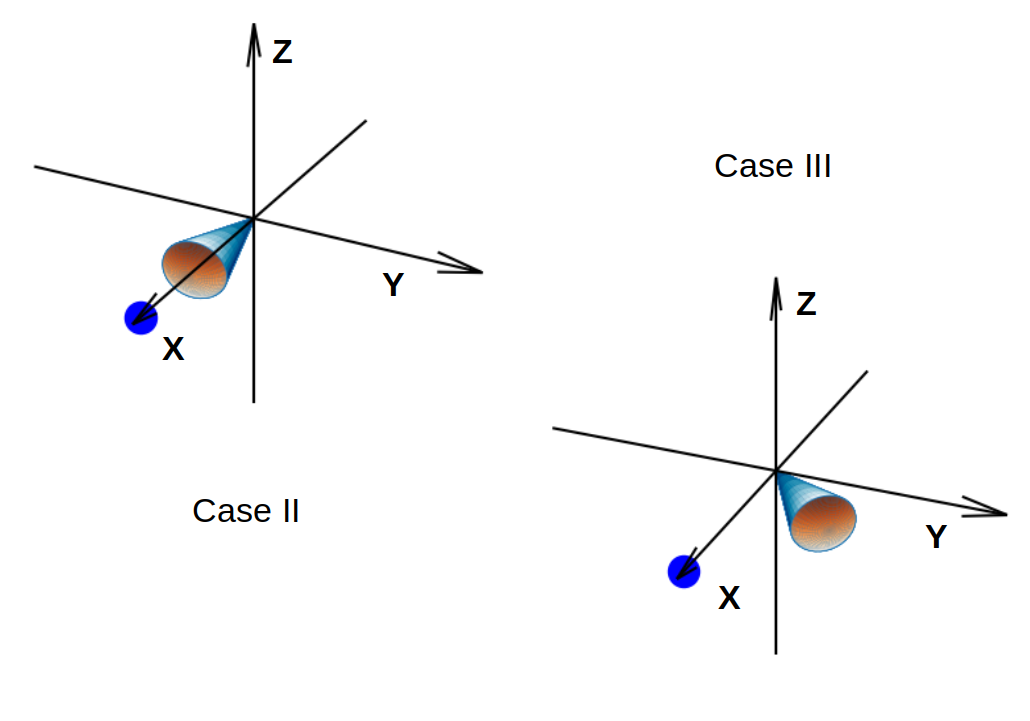}
    \caption{ Left: Schematic plot showing the CME configuration for the base case. Plasma downstream of the shock is divided into parcels. These parcels are used to track the ENA production and propagation to the observer. The observer is along the X axis at 1 au and the CME propagates along the Y axis.  Right: Another two cases, case II and case III are also considered.  In case II, the CME propagates toward the observer; and in case III, the CME propagates $45^{\circ}$ off from the $+Y$ direction. } \label{fig:downstreamParcels}
\end{figure}

Here we reexamine the case considered in \citep{Wang2014ApJ...793L..37W} and include another two cases with different CME propagation directions to obtain an estimate of ENA flux range at 1 au. Our treatment is similar to our previous work \citep{Wang2014ApJ...793L..37W} but with 
a few differences.  As in \citep{Wang2014ApJ...793L..37W}, we assume protons are accelerated at the shock and then distributed uniformly downstream of the shock. This is based on the DSA mechanism and has been adopted in 
our previous large SEP event simulations \citep{Li2003,Li2005,Li2012,Li2021}. 
Since the turbulence downstream of the shock is a lot stronger than that upstream of the shock (see e.g. \citep{Lee1983,Zank2000,Li2003}), accelerated particles can be kept downstream of the shock for a long period of time.
In \citep{Wang2014ApJ...793L..37W}, we assumed there is no leakage of accelerated particles from downstream of the shock. This was mostly for simplicity since accelerated particles can precipitate back to the sun along open field lines. Indeed, \citet{JinMeng2018ApJ...867..122J} has explored the possibility that the long duration gamma ray events are due to shock acceleration protons. In such a scenario, accelerated protons downstream of the shock can steadily precipitate to the solar surface.  Therefore, in this work, we include a decay of the accelerated protons downstream of the shock.  As an estimate of the decay time, we  refer to \citet{Li.etal2012Twin}, who, from a statistical study of twin-CME events,  suggested that a decay time scale of the turbulence in large SEP events is around $9$-$13$ hours. We use a decay time $\tau=10$ hours in this work. We also set our inner boundary at $r=1.02 R_s$, which differs from that used in \citep{Wang2014ApJ...793L..37W}, $1.5 R_s$. We further improve the treatment of ENA propagation from downstream of the shock to the observer.
In \citep{Wang2014ApJ...793L..37W}, downstream medium was divided into shells and ENAs produced in individual shells are assumed to propagate to the observer all from the shell center. This is refined in our current work.   
We now divide the downstream region of the shock into multiple parcels, as shown in the left panel of Figure~\ref{fig:downstreamParcels}. ENAs are produced and followed in individual  parcels. Since ENAs in different parcels propagate to the observer along different paths, our current treatment will lead to a more accurate survival probability computation. 
Finally, a correction factor $cos(\theta)$ to the flux expression,   equation~(4) in \citep{Wang2014ApJ...793L..37W} is included, see equation~(\ref{eq:currentJ}).

Figure~\ref{fig:downstreamParcels} shows the configuration of the ENA production process for the CME shock case. The left panel depicts the base case:  the observer locates at 1 au along the $X$ axis and the CME is propagating to the right along the $+ Y$ direction, i.e., $\phi =90^\circ$ where $\phi$ is the angle between the sun-observer line and the CME propagation direction. The plasma downstream of the shock is divided into multiple parcels. ENA production is followed in these parcels. ENAs produced in these parcels can propagate along straight lines to the observer. These trajectories differ for different parcels, and lead to different survival probabilities. Right panel of Figure~\ref{fig:downstreamParcels} shows two other cases with different 
CME propagation directions. In case II, the CME propagates toward the observer with $\phi=0^{\circ}$. In case III, the CME propagates $45^{\circ}$ off from the $+X$ and $+Y$ directions, i.e. $\phi=45^\circ$.
For our simulation, the  shock has a constant speed of $V_{sh}=1500$ km/s and a constant compression ratio of $s=3.5$. The open angle of the shock is $60^{\circ}$ and the shock is followed up to $30 R_s$. As in the flare ENA case, we use the Leblanc model \citep{Leblanc1998SoPh..183..165L} to compute the solar wind density.

\begin{figure}[ht]
    \centering
   \includegraphics[width=0.48\textwidth]{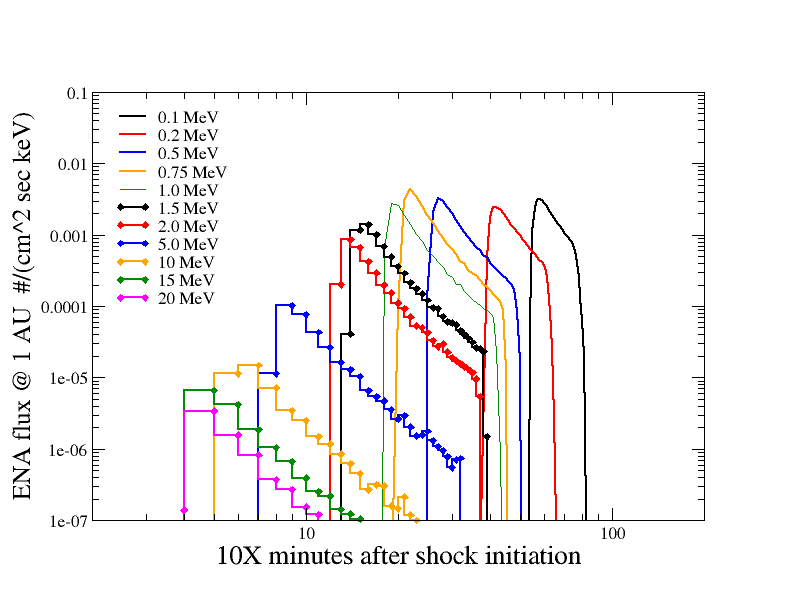}
    \includegraphics[width=0.48\textwidth]{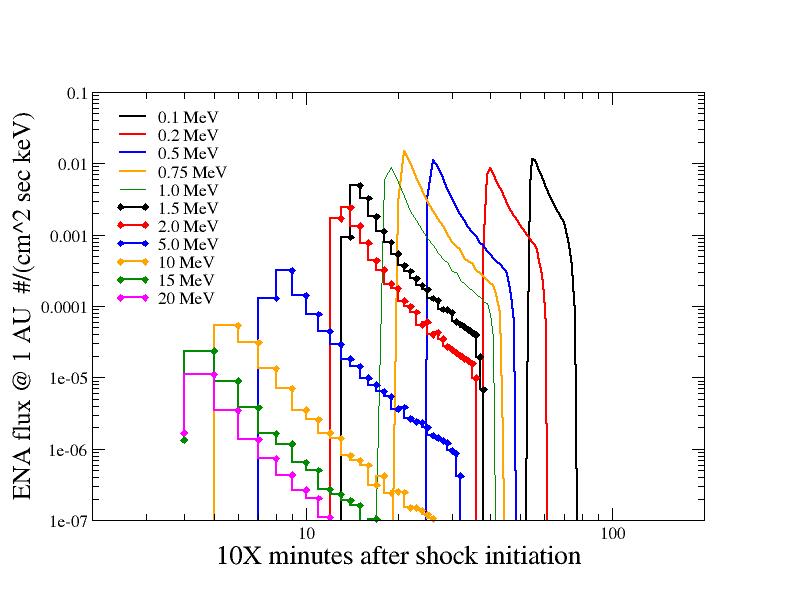}
     \includegraphics[width=0.48\textwidth]{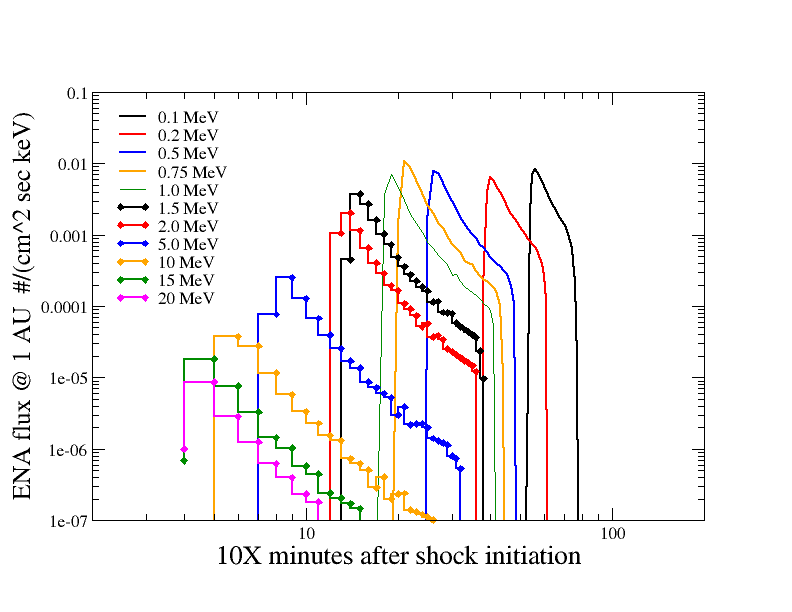}
   \includegraphics[width=0.48\textwidth]{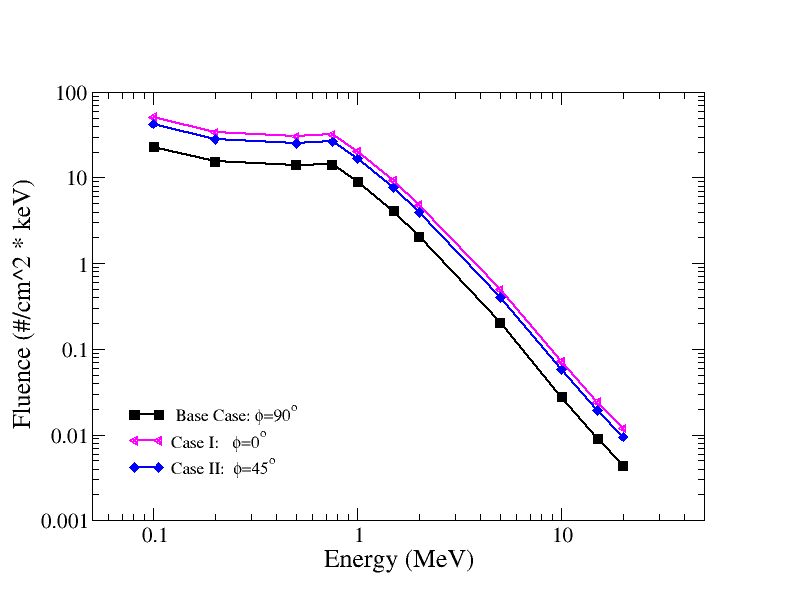}
   \caption{Upper left, upper right and lower left panels  show time Profiles of ENA hydrogens produced at CME-driven shocks for an observer at 1 au, for the base case, case II and case III, respectively. Eleven energies are considered.  Shocks is followed up to $30 R_s$. Lower Right panel is the fluence of the ENA hydrogen for the time duration shown in the other three panels. Note the bent-over at 1 MeV for the ENAs. The parent energetic protons has a power law extended to $0.02$ MeV. The bent-over is due to the energy dependence of the various charge cross sections shown in the Appendix~\ref{sec:ProductionLoss}.      } \label{fig:CME_ENA_TimeProfile_Fluence}
\end{figure}

Figure~\ref{fig:CME_ENA_TimeProfile_Fluence} plots the time profiles and the fluence of ENAs for the three cases shown in Figure~\ref{fig:downstreamParcels}. The upper left, upper right, and lower left panels show the time profiles for the base case, case II and case III, respectively.  For all three cases, ENAs of $11$ energies are considered.  The three time profiles are similar. Consider the base case (upper left panel). The x-axis is the time after shock initiation, in unit of $10$ minutes; 
and the y-axis is the ENA flux at the observer, in unit of \#/(cm$^2 \cdot$ sec $\cdot$ keV). The observer first see  
the $20$ MeV ENAs arriving  $\sim 40$ minutes after the shock initiation. The flux can reach {$7*10^{-3}$ cm$^2 \cdot$ sec $\cdot$ keV}. 
It then decreases, reflecting the fact that the density of energetic protons  decreases with time as the shock propagates out. As the ENA energy become smaller, their first arrival times become later and the flux increases with decreasing energy, till $E=0.75$ MeV. Below $E=0.75$ MeV the flux shows a more plateau feature and drops slightly. This behaviour is due to the energy dependence of the charge exchange cross sections that are responsible for the ENA production. See Figure~\ref{fig:crossSections} in Appendix~\ref{sec:ProductionLoss}. Comparing to the base case, cases II and III are comparable and show larger fluxes than the base case. This is easily understood from Figure~\ref{fig:downstreamParcels} because the ENAs produced in these two cases travel shorter distances and through less dense solar atmosphere to the observer and consequently have larger survival probabilities. The lower right panel of Figure~\ref{fig:CME_ENA_TimeProfile_Fluence} plots the fluence  for these three cases. Note the relatively plateau-like behavior below $E=0.75$ MeV, which is the consequence of the energy dependence of the relevant charge exchange cross section. 
Above $1$ MeV, the ENA fluence spectrum shown here is comparable to that inferred in \citet{Mewaldt2009ApJ...693L..11M}. 
The general shape of the CME shock ENA fluence is similar to the parent energetic ion spectrum which is a power law. This is in stark contrast to the flare ENA case (see next section) where the ENA fluence does not resemble the parent energetic ion spectra.

\section{ENAs from solar flares} \label{sec:FlareENAs}

We examine ENA production by solar flares in this section. Both electrons and ions are efficiently accelerated at solar flares, and the accelerated electrons and ions lead to the emission of hard X-rays and gamma rays.
It is generally accepted that 
acceleration may occur at reconnection current sheets and/or by turbulence in the flare loops. 
Observations of hard X-ray and gamma rays suggest that the accelerated electron and ion spectra can be approximated by a power law.
Power law like spectra are supported by earlier theoretical works by \citep{Miller1995ApJ...452..912M, Petrosian2012SSRv..173..535P}, where 
ions are accelerated in flare loops by MHD turbulence via second order Fermi acceleration. More recent PIC simulations of ions in flare reconnection site also found a power law spectrum  \citep{Zhang2021PhysRevLett.127.185101}.

\begin{figure}[ht]
    \centering
   \includegraphics[width=0.45\textwidth]{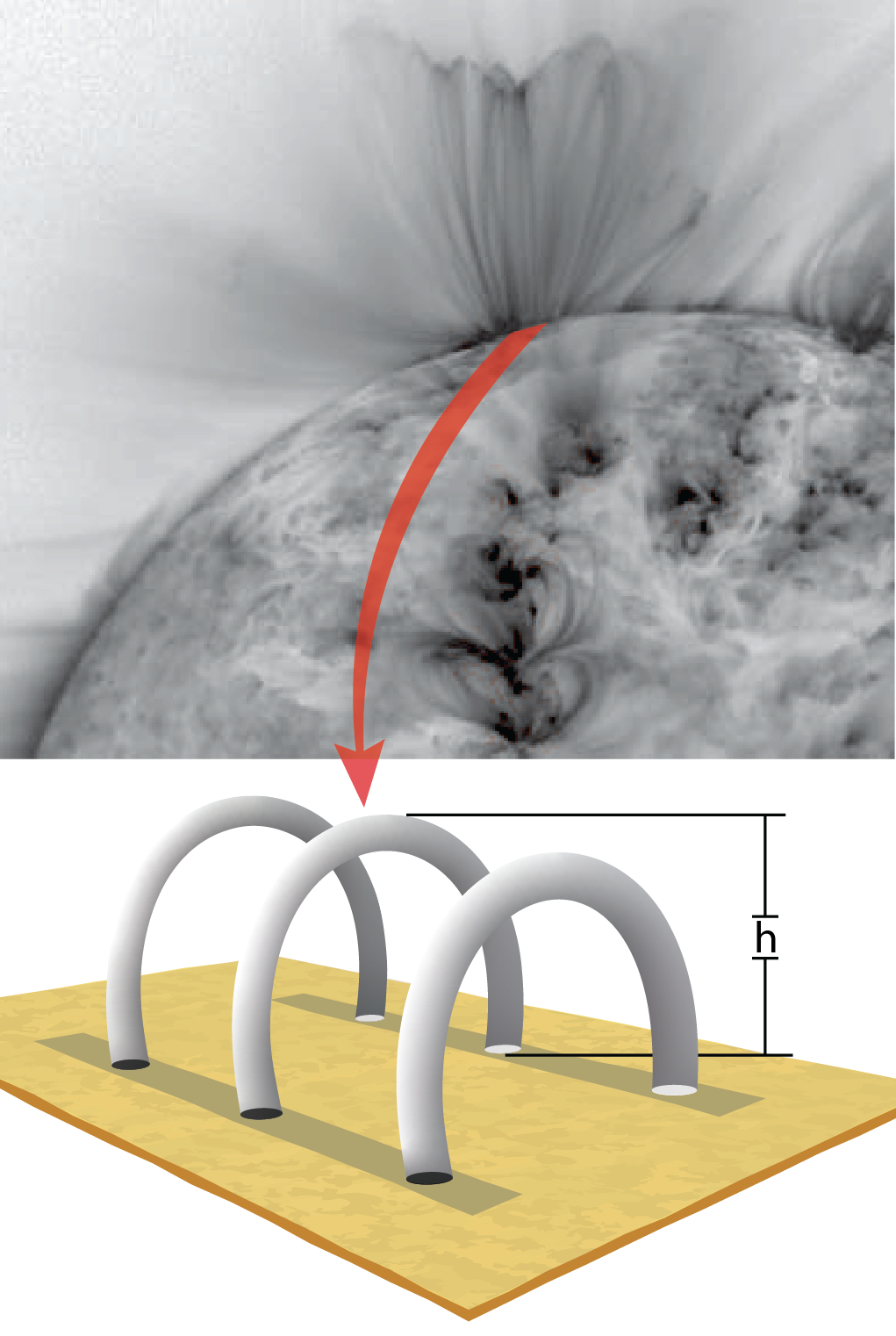}
    \caption{Cartoon showing the postflare loops in solar flares. Large scale and high postflare loops are potential production site of solar ENAs. Upper image is adopted from \citep{West2015ApJ...801L...6W}.} \label{fig:FlareLoops}
\end{figure}

Once accelerated, ions precipitate down to the solar surface along post-flare loops.  The density of a post-flare loop can be constrained by free–free continuum emission for hot loops. In a recent work, \citet{Jejcic2018ApJ...867..134J} reported an electron density as high as $10^{13}$ cm$^{-3}$, $10$ to $100$ higher than that at typical flare loops. 
At a density of $\sim 10^{11}$ cm$^{-3}$, ENAs can be easily produced in these loops. 
Once produced, ENAs are not constrained in the loops and can propagate in all directions. 
However, if the loops are low, the solar atmosphere density in the surrounding environment can be too dense to allow these ENAs to escape from the Sun. Therefore to observe flare ENAs, the flare loops must be high.  
The height of flare loops can be estimated from the looptop hard X-ray observations.
A recent study of looptop hard X-ray source of solar flares \citep{Effenberger2017ApJ...835..124E} showed that the height of a typical flare ranges from $10$ to $50$ Mm. 
If ions are accelerated at and below this height at flares, no ENAs can survive as they propagate out.  However, using the Sun Watcher with Active Pixels (SWAP) EUV imaging solar telescope, \citet{West2015ApJ...801L...6W} examined an M2.2 flare which occurred on 2014 October 14 and found that the post-flare loops were long-lasting, and reached a height of over $400$ Mm ($\ge 0.5\ {{R}_{\odot }}$) $\sim 48$ hours after the eruption.
\citet{West2015ApJ...801L...6W} 
argued that the giant arches in this event are similar to ordinary post-flare loops and are the results of a long-lasting ($48$ hours) magnetic reconnection occurred along a large-scale current sheet \citep{Forbes2000JASTP..62.1499F}.  This continuous magnetic reconnection provides the energy source to heat the loop and can accelerate particles. Besides magnetic reconnection, turbulence inside the loop can also lead to stochastic acceleration of ions \citep{Ryan2000SSRv...93..581R}. We note that the magnetic reconnection at the current sheet and the enhanced turbulence inside the large postflare loop may be intimately related. \gangli{In a recent work by \citet{Cheng2018ApJ...866...64C}, the authors examined the 2017 09 10 flare and showed that a Kolmogorov-like turbulence spectrum  can develop in the current sheet above the flare loops. Presence of such a turbulence implies that particles can be accelerated in the turbulent current sheet in a similar way as in flare loops through a second order Fermi acceleration process \citep{Miller1995ApJ...452..912M, Petrosian2012SSRv..173..535P}.
The spatial extension of the current sheet 
is similar to the flare loops that are beneath it  \citep{Cheng2018ApJ...866...64C,French2019ApJ...887L..34F}, but the density of the current sheet is,  however, smaller than 
the density in the postflare loops. 
Indeed \citet{French2019ApJ...887L..34F} concludes that the density in the current sheet is $\sim < 10^{10}$/cm$^3$. This is $100$ times smaller than the 
density inferred in the lower flare loop as reported in \citep{Jejcic2018ApJ...867..134J}, and 
is $10$ times smaller than what we assume for the post flare loops, $10^{11}$/cm$^3$. The ENA production in these 
current sheets is therefore much smaller than
in postflare loops. So we only consider ENA producion in postflare loops in this work.
However, we remark that these turbulent current sheets can be potential sites of ENA production, and if future ENA probes have high enough sensitivities, it is possible to obtain direct observations of these current sheets through ENA observations.} A cartoon showing these post flare loops are shown in Figure~\ref{fig:FlareLoops}.

Continuous acceleration, as suggested by \citet{Ryan2000SSRv...93..581R}, has been identified as a possible scenario for the long duration gamma ray events
\citet{deNolfo2019ApJ...879...90D}.
Long duration gamma ray events are not uncommon. Recently \citet{Share018ApJ...869..182S} examined $\sim 30$ long duration gamma ray events and found 
that the energy spectral indices of $>300$ MeV proton producing gamma rays range from $2.5$ to $6.5$, similar to  typical flare events. In a recent study, \citet{deNolfo2019ApJ...879...90D} compared the gamma-ray-producing proton numbers with the in-situ SEP proton numbers in long duration gamma ray flares and found a poor correlation. Their study supports the continuous acceleration in the post-flare loop scenario, as suggested by \citet{Ryan2000SSRv...93..581R}. We point out that the event reported in \citep{West2015ApJ...801L...6W}, despite having large post-flare loops, was not a 
long duration gamma ray event. This is possible if particles are not accelerated to high enough energies ($\sim 100$ MeV/nuc) to produce gamma rays.

We now examine ENAs from post-flare loops. 
We model the post-flare loops as semi-circle tubes.
We assume that the loop has a height (radius) of $h(t)$, which increases with time. We assume the starting height of the post flare loop is $0.04 R_s$ ($\sim 28 $Mm), and 
a rising rate of $V_r=3$ km/s \citep{West2015ApJ...801L...6W}. This gives a height of $H=0.22$, $0.41$, and $0.60$ $R_s$ when $t=$ 12, 24, and 36 hours, respectively.  
The cross section of the tube can be assumed to be a circle with a radius as $a$. One can take $a$ to be $\sim 700$ km, which is comparable to the half width for a typical flare ribbon. However, as we will see below, the ENA production depends on the 
total number of accelerated protons and does not depend on the choice of $a$ and the number of loops we consider.

We also assume a constant proton density inside the flare loop.  By way of example, we assume a loop density of $10^{11}$ cm$^{-3}$.
This is smaller than that obtained in 
\citep{Jejcic2018ApJ...867..134J}, but larger than the density at the solar surface, which is $\sim 10^{9-10}$ cm$^{-3}$.
As a simplification, we assume the acceleration process \citep{Ryan2000SSRv...93..581R} is time independent 
and the production rate of energetic protons, $\alpha$, is a constant during the rising phase of the post-flare loop. We denote the duration of the rising phase to be $T$, and the total number of accelerated particle $N_0= \alpha T$.  
Once accelerated these particles can precipitate to the solar surface. We model this as a loss process with an energy-independent decay time $\tau$.  The total number of accelerated particles $N(t)$ in the loop is given by,
\begin{equation}
\frac{d N(t)}{dt} = \frac{N_0}{T}\theta(T-t) - \frac{N(t)}{\tau} 
    \label{eq:productionI}
\end{equation}
where $\theta(t)$ is the Heaviside function. The solution of 
equation~(\ref{eq:productionI}) is, 
\begin{equation}
    N(t) = N_0 \frac{\tau}{T} \left [ (1-e^{-t/\tau}) *\theta(T-t)
+ (1-e^{-T/\tau})e^{-(t-T)/\tau} *\theta(t-T) \right ].  \label{eq:Nt}
\end{equation}
In equation~(\ref{eq:Nt}), $N_0$ can be constrained from the following consideration. In the long duration gamma ray events examined by \citep{Share018ApJ...869..182S}, the authors inferred that accelerated particles 
 at high energies ($>$300 MeV) in the loops is about $0.01$ to $0.5$ of that of the accompanying SEP events, presumably accelerated at the CME-driven shocks. Assuming this ratio is energy independent, then one can estimate the range of $N_0$ from the CME-driven shock case.  Alternatively, one can estimate $N_0$ from an energy budget point of view.
 In a study of the CME/Flare Energy Budget for two  \citep{Emslie2005JGRA..11011103E},
 and subsequently for $38$  large SEP events,
 \citet{Emslie2012ApJ...759...71E}
 found that the energy budget for $> 1$ MeV flare ions can reach $ \epsilon \sim 4*10^{31}$-$10^{32}$ erg, which can be comparable and even larger than those observed in-situ. 
In this work, we estimate $N_0$ by assuming the total energy for the accelerated particles ($> 1$ MeV) is $\epsilon = 10^{31}$ erg. With a 
 source spectrum of the accelerated protons given by,  
\begin{equation}
    f(E,t) = \frac{N(t) (\gamma_1-1)}{E_0}  \left [ ( \frac{E}{E_0})^{-\gamma_1} \theta(E_b-E)
     + (E_b/E_0)^{-\gamma_1} (\frac{E}{E_b})^{-\gamma_2}  \theta( E-E_b ) \right ] 
    \quad 
    \label{eq: distrFunc}
\end{equation}
where $E_0$ is the injection energy, $E_b$ is the break energy, $\gamma_1=2.5$ is the spectral index at energies below $E_b$ and $\gamma_2=5.5$
is the spectral index at energies above $E_b$. This gives, 
\begin{equation}
    N_0 \approx (\frac{\gamma_1-2}{\gamma_1-1}) \left ( \frac{\epsilon E_0^{1-\gamma_1}} {1-(E_b)^{2-\gamma_1}} \right )
    \label{eq:N0}
\end{equation}
For a choice of $E_0 = 0.02$ MeV, $E_b = 30$ MeV, $\gamma_1=2.5$ and $\gamma_2=5.5$, we find  $N_0=9 * 10^{38}$. Equation~(\ref{eq: distrFunc}), together with equations~(\ref{eq:Nt}) and ~(\ref{eq:N0}) describe the energetic proton source, as a function of time, for the ENAs inside the post-flare loop. 

One can now compute the production of ENAs and obtain the time profiles and fluence of ENAs as observed at 1 au. We consider three cases with $T=12$, $24$, and $36$ hrs, corresponding to a final loop height of $H = 0.22$, $0.41$, and $0.60$ R$_s$, respectively. 
In all cases  $\tau = 3$ hr. We further assume that the flare locates at $\phi=0$ degree, i.e. in a face-on situation. For other viewing angles, the results are qualitatively similar.

\begin{figure}[ht]
    \centering
   \includegraphics[width=0.48\textwidth]{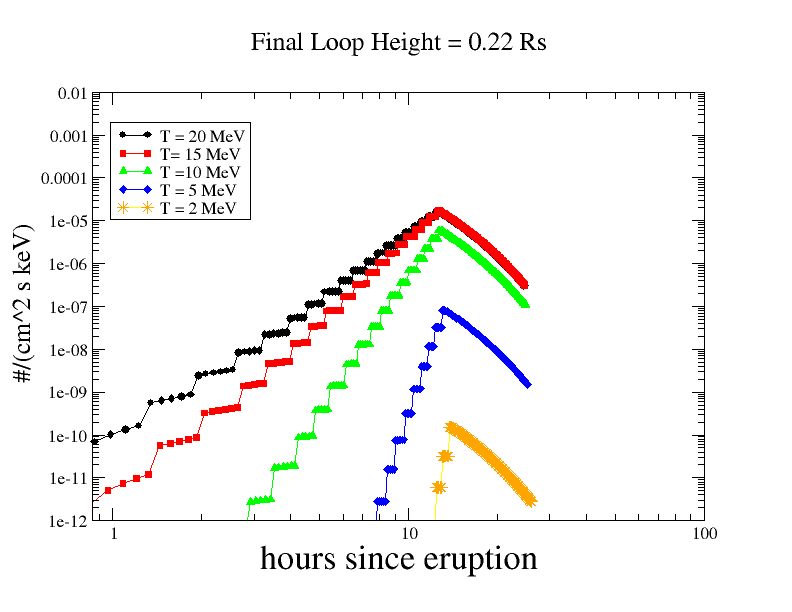}
   \includegraphics[width=0.48\textwidth]{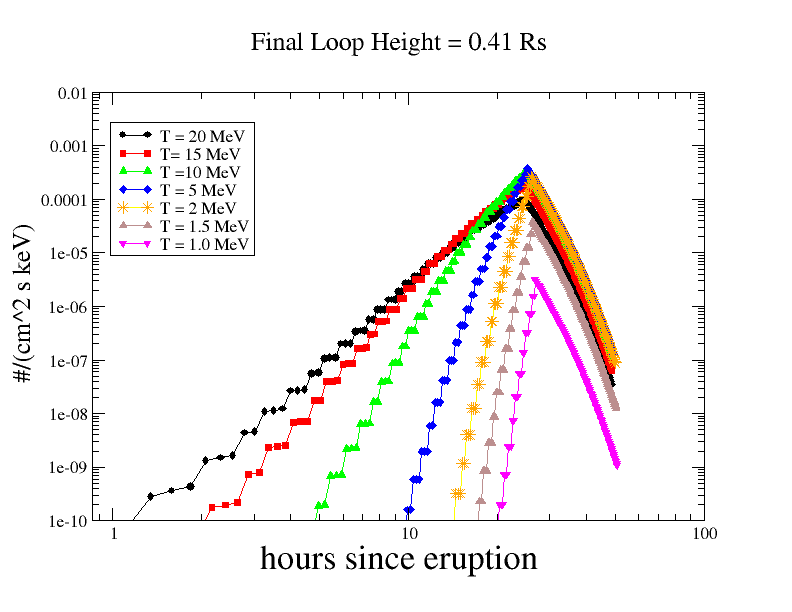}
      \includegraphics[width=0.48\textwidth]{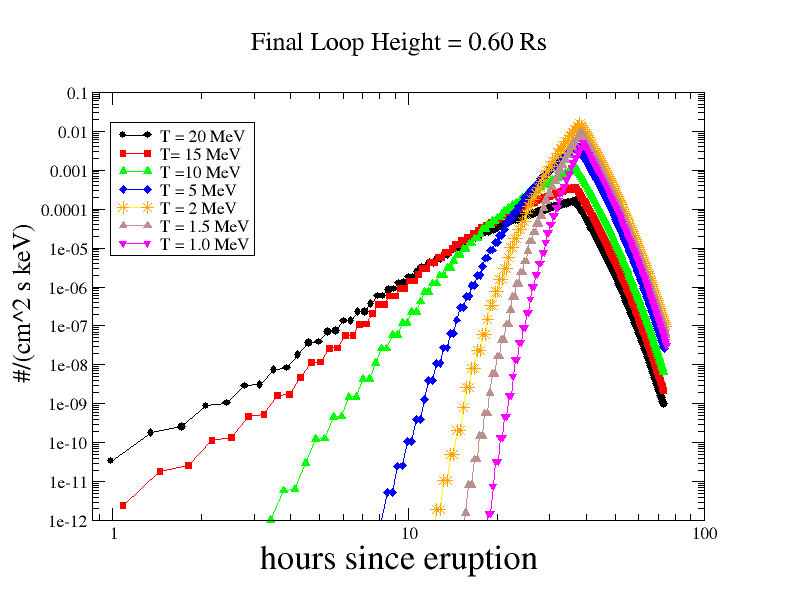}
   \includegraphics[width=0.48\textwidth]{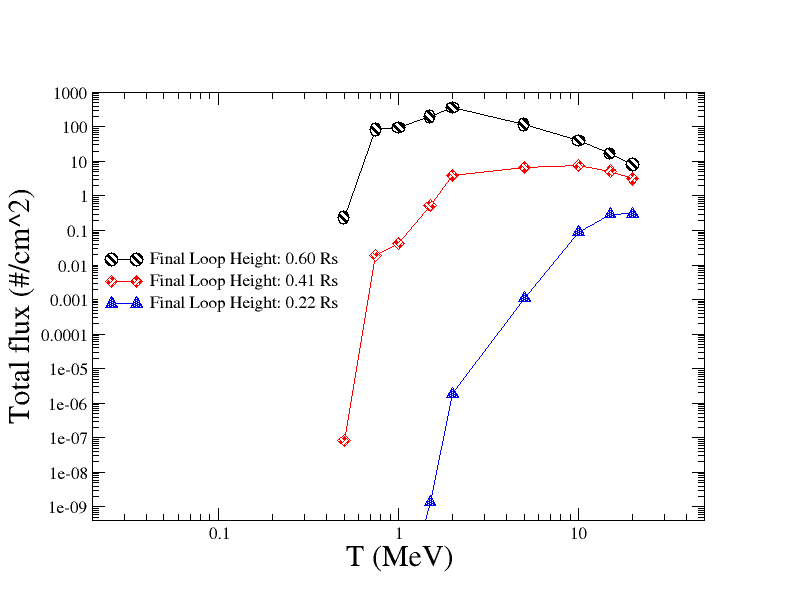}

    \caption{Upper left, upper right, and lower left: time profiles of solar flare ENAs for a loop with a final loop height $h=0.22 R_s$, $0.41 R_s$, and $0.60 R_s$, respectively. Lower Right: total ENA fluence for the three cases considered. See text for details.
     } \label{fig:flareENAs}
\end{figure}

Figure~\ref{fig:flareENAs} plots the time profiles and fluence of the flare ENAs.  The upper left, upper right and lower left panels are time profiles 
for the three choices of the final flare loop heights. 
 Seven energies are considered. These are $1.0$, $1.5$, $2$, $5$, $10$, $15$, and $20$ MeVs. 
 As can be seen from these panels, high energy ENAs arrive earlier due to a short propagation time from the Sun to 1 au.  In all three panels, the peak of the time profiles occur shortly after the loops reach the maximum height. The energy dependence of the peak intensity (and the fluence, see the lower right panel) strongly depends on the loop height. If the loop height is $0.22 R_s$ (upper left panel), the peak intensity of $T=2$ MeV ENAs is 
 $5$ orders of magnitude smaller than 
 that of the $T=15$ MeV ENAs. Furthermore, there is no ENAs with $T< 2$ MeV. In comparison, when the loop height is $0.41$ or $0.6$ R$_s$,  the peak intensity of $T=2$ MeV ENAs is similar to that of the $T=15$ MeV ENAs. This energy dependence can be also seen from the fluence plot shown in the lower right panel  of Figure~\ref{fig:flareENAs}.  
 When the loop height is $0.60 R_s$, the fluence has a maximum $\sim 800$/cm$^2$ at $T=2$ MeV, and at $T= 20$ MeV, the fluence is about  $10$. When the loop height is $0.22 R_s$, however, the fluence of 2 MeV ENAs drop by a factor of $4*10^8$ to $\sim 2*10^{-6}$/cm$^2$. In comparison, the flunece of $20$ MeV ENA drops only by a factor of $50$, to $0.2$/cm$^2$. This big difference of ENA fluence at 1 au for different flare loop height is due to efficient loss of ENA close to the Sun. Although plenty of ENAs are produced in the flare loop, they can not escape the high density solar atmosphere if the flare loop is not high enough. Note that during the eruption phase of solar flares, the height of flare loops, as seen from X-ray imaging, is a lot smaller 
 than $0.22R_s$ \citep{Effenberger2017ApJ...835..124E},
 therefore we expect no ENAs during the eruption phase of solar flares. However, large post flare loops, as those 
 reported in \citep{West2015ApJ...801L...6W}, can reach $0.5 R_s$. Our calculations show that there will be clear ENA signals from such a flare. We do note that the absolute amplitude and the shape of the ENA fluence depend on the solar atmosphere density model as well as the relevant charge exchange cross sections (see Appendix \ref{sec:ProductionLoss}).  Nevertheless, because the ENA fluence, and in particular, its energy dependence, sensitively depend on the flare loop height, one can use the ENA fluence as a probe of the flare loop height. We point out that these large post flare loops may not be common. Consequently, flare ENAs may not be common either.  Note that both the time profiles and the fluence for flare ENAs shown in Figure~\ref{fig:flareENAs} are vastly different from their counterparts in shock accelerated ENAs shown in  Figure~\ref{fig:CME_ENA_TimeProfile_Fluence}. This suggests that one can use ENA observations to discern if the parent energetic ions are accelerated at CME-driven shock or at solar flares.

\section{Conclusions} \label{sec:Conclusions}

Understanding the underlying particle acceleration process in large SEP events has been one of the central problems in heliophysics research. With only in-situ observations of energetic ions, questions such as the relative roles of magnetic reconnection in flares vs shock acceleration at CME shocks, and how to discern the effects of acceleration from that of transport, can be very hard to answer. In part, this is because our basic understanding of the near-Sun conditions and the physical processes involved in the production of SEP events is hampered by our inability to make direct measurements
near the acceleration sites and to remove the effects of transport. ENA observations can significantly advance our understanding of SEP acceleration at its source because ENAs do not interact with IMF and is not affected by the transport  effect.

In this paper, we examine the production of ENAs at CME-driven shock fronts and in solar flares. We compute the time profiles and fluence of ENAs for these two scenarios. Our calculations suggest that in large SEP events where ions are efficiently accelerated at CME-driven shocks, ENAs are copiously produced behind the shock. At 1 au the flux of these ENAs are at a level that can be readily measured by a dedicated ENA detector. ENAs can also be produced in flares where large scale and high postflare loops exist.  The time profiles and fluence of ENAs for these two scenarios differ considerably. This offers us an opportunity to constrain the underlying particle acceleration process via ENA observations. Our work also forms a theoretical basis for interpreting future ENA observations.

\begin{acknowledgments}
This work is supported in part by NASA grants 80NSSC19K0075  and 80NSSC20K1783, and NSF grant 2149771 at UAH. Work at SwRI is partially supported by NASA LWS grants 80NSSC19K0079 and 80NSSC20K1815. And work at APL is partially supported by NASA contract 80MSFC19F0002.
\end{acknowledgments}

%





\appendix

\section{Production and Loss of solar ENAs} \label{sec:ProductionLoss}
\textbf{Production:}
We examine the ENA production at solar flares and CME-driven shocks in this work. 
The underlying ENA production process is the same for both cases and is through charge exchange reactions.  At time $t$ and location ${\bf r}$, the production rate of ENA is, 
\begin{equation}
     A( {\bf r}, E,t) = \frac{dn}{dt dE} = \sum_i n_i \cdot \sigma_i \cdot v \cdot f({\bf r}, E) \label{eq:productionA} 
\end{equation}
   Here $f({\bf r}, E)$ is the distribution function of the 
   accelerated proton from either the CME-driven shock or the flare site; $E=\frac{1}{2}m_p v^2$ is the kinetic energy of the energetic proton and we consider non-relativistic case; 
   the sum is for all contributing charge exchange processes. For the case of solar composition, the following three charge-exchange interactions are the most relevant:
\begin{equation}
p + O^{6+} \rightarrow H + O^{7+}, \quad
p + C^{4+} \rightarrow H + C^{5+}  \quad 
p + H \rightarrow H + p,
\end{equation}
The abundance ratio of O$^{6+}$/p is $\sim 10^{-3}$, and C$^{4+}$/O$^{6+}$ is $\sim 0.067$
\citep{vonSteiger2000JGR...10527217V}. For neutral hydrogen, ionization by impact collision and EUV balance the recombination and charge exchange collisions, leading to a ratio of neutral H to proton to be $\sim 2.6*10^{-7}$ \citep{Damicis2007JGRA..112.6110D}. 

\begin{figure}[ht]
    \centering
   \includegraphics[width=0.75\textwidth]{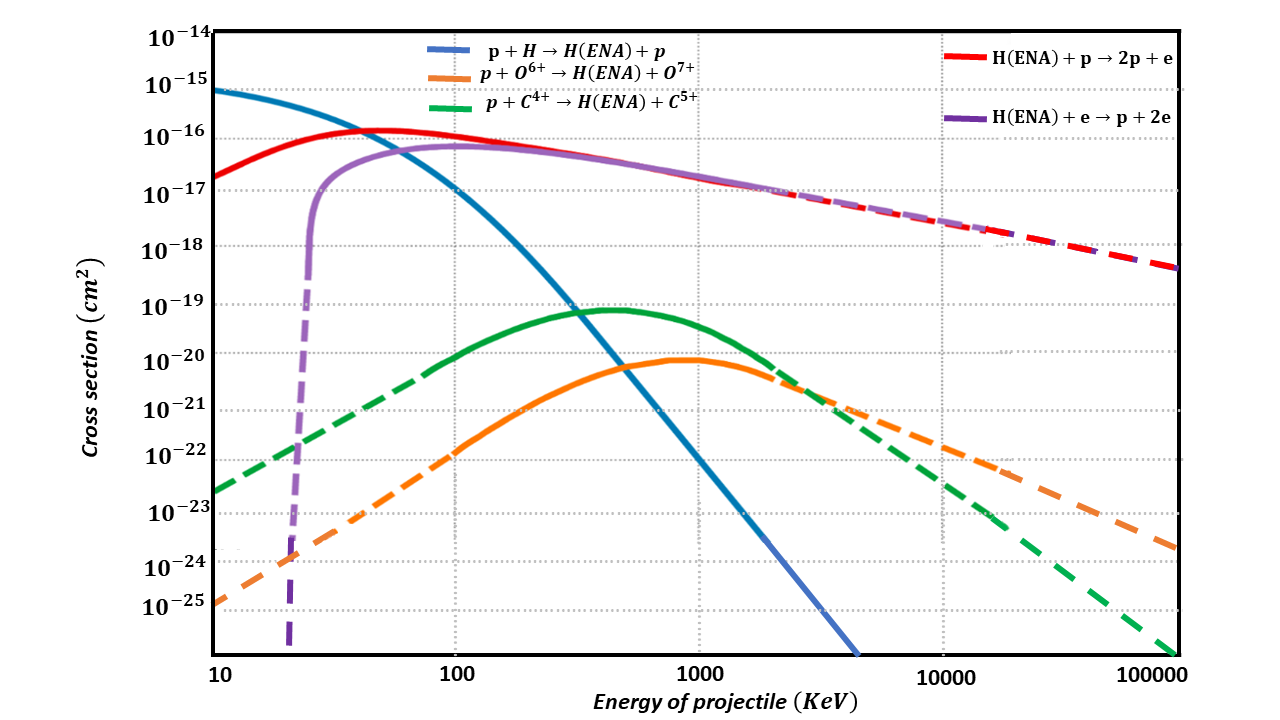}
    \caption{The relevant cross sections for ENA production and loss. Adopted from \citep{Wang2014ApJ...793L..37W, Wang2022E&PP....6...42W}. Dashed lines signal extrapolations.  } \label{fig:crossSections}
\end{figure}

The corresponding cross sections for the three charge-exchange interactions, as a function of proton energy, are shown in Figure~\ref{fig:crossSections}. These cross sections were obtained from theoretical calculations \citep{Gruntman2001JGR...10615767G, Kuang1992JPhB...25..199R} and are subject to uncertainties. The energy range for these cross sections are also limited. Following \citet{Wang2022E&PP....6...42W}, we have extended them to a larger energy range whenever necessary.
Note that as in \citep{Wang2014ApJ...793L..37W}, we ignore charge exchanges by other ions 
(He$^+$, N$^{5+}$, etc) due to their smaller abundances.  Including these would marginally increase the 
ENA production rate.

\textbf{Propagation and Loss of ENAs:} Once produced, solar hydrogen ENAs leave their birth places along ballistic trajectory, subject to losses due to primarily impact ionization and EUV ionization. The cross sections for the two most important impact ionization processes are also shown in Figure~\ref{fig:crossSections}.
The differential flux 
$J({\bf r}, v {\hat n}, t)$ (with unit of s$^{-1}$ cm$^{-2}$ keV$^{-1}$), at location ${\bf r}$, time $t$, and along the direction of ${\hat n}$, is given by,

\begin{eqnarray}
J({\bf r}, v{\hat n}, t)
=\int_0^{t} dt'\int d^3 {\bf v'} d^3 {\bf r'}
\frac{A({\bf r'}, {\bf v'}, t') h({\bf r}-{\bf r'}, v)}
{4 \pi |{\bf r}-{\bf r'}|^2}
\delta(v-v')
\delta( t-t'-\frac{|{\bf r}-{\bf r'}|}{v})cos(\theta) \label{eq:currentJ}
\end{eqnarray}
where $cos(\theta) = ({\bf r}-{\bf r'}) \cdot \hat{\bf n}/|{\bf r}-{\bf r'}|$ and $h({\bf r}-{\bf r'}, v)$ is the survival probability of the neutral hydrogen at location ${\bf r}$, produced at ${\bf r'}$. The survival probability $h({\bf r}-{\bf r'}, v)$ depends on the travel history and its speed $v$ of the ENA hydrogen and is computed by \citep{Wang2014ApJ...793L..37W}, 
\begin{equation}
    h({\bf r}-{\bf r'}, v) =exp(-\int_0^{|{\bf r}-{\bf r'}|} \gamma({\bf r'}) {dl} )
\end{equation}
where the integration $dl$ is along the direction ${\bf r}-{\bf r'}$ and $\gamma$ is the total loss rate. We consider three loss processes here: electron impact ionization, proton impact ionization, and photo-ionization.  The loss rate for these processes are \citep{Wang2014ApJ...793L..37W},
\begin{equation}
    \gamma_{eH} = {\rho_{sw,e}({\bf r}) \sigma_{eH}}, \quad 
    \gamma_{pH} = {\rho_{sw,p}({\bf r}) \sigma_{pH}}, \quad 
    \gamma_{\gamma H} = 4 * 10^{-3} (\frac{r_s}{r})^2 \frac{1}{v}.
 \end{equation}
 
For both the flare ENAs and the CME-shock ENAs, the treatment of ENA production and propagation/loss is the same. The difference between them is the region of the energetic ion source. Comparing to the 
CME case,  the post-flare loops is more localized.


%



\end{document}